\documentclass[aip,jcp,reprint,groupedaddress]{revtex4-1}

\usepackage{amsmath}
\usepackage{txfonts}

\usepackage[T1]{fontenc}
\usepackage{textcomp}

\usepackage{bm}
\bmdefine{\bVector}{b}
\bmdefine{\BVector}{B}
\bmdefine{\eVector}{e}
\bmdefine{\EVector}{E}
\bmdefine{\fVector}{f}
\bmdefine{\FVector}{F}
\bmdefine{\gVector}{g}
\bmdefine{\pVector}{p}
\bmdefine{\PVector}{P}
\bmdefine{\qVector}{q}
\bmdefine{\QVector}{Q}
\bmdefine{\rVector}{r}
\bmdefine{\RVector}{R}
\bmdefine{\sVector}{s}
\bmdefine{\uVector}{u}
\bmdefine{\vVector}{v}
\bmdefine{\VVector}{V}
\bmdefine{\muVector}{\mu}
\bmdefine{\OmegaVector}{\Omega}

\usepackage[pdftex]{graphicx}

\usepackage{enumerate}

\draft 

\begin{document}


\title{Particle Monte Carlo simulation of string-like colloidal assembly in 3 dimensions} 



\author{Yuki Norizoe}

\author{Toshihiro Kawakatsu}
\affiliation{Department of Physics, Tohoku University, 980-8578 Sendai, Japan}


\date{1 December, 2011}

\begin{abstract}
As an extension of the former study on 2-dimensional systems, we simulate phase behavior of polymer-grafted colloidal particles in 3 dimensions by molecular Monte Carlo technique in the canonical ensemble. We use a spherically symmetric square-step repulsive interaction potential, which has been obtained using self-consistent field calculation. In previous articles, we have studied these model colloids in 2 dimensions and found that these particles, although their interaction is purely repulsive, self-assemble into a string-like assembly, in a narrow region of physical parameter sets. In the present work, we show the existence of the string-like assembly in 3-dimensional systems and study the statistical properties of the arrangement of these strings. The average string length diverges around the region where the melting transition line and the percolation transition line cross, which has also been found in 2 dimensions.
\end{abstract}


\maketitle 



%
%

%

\section{Introduction}
\label{sec:Introduction}
We studied, in our previous works, the phase behavior of a polymer-grafted colloidal system in 2 dimensions~\cite{Norizoe:2005,2010:NorizoeMuPTLetterArXiv,2011:NorizoeMuPTFullArXiv,2011:NorizoeFrogNVT2DArXiv,Norizoe:2003April,Norizoe:2003November,Master'sThesis}. In these studies, linear diblock copolymers were grafted onto the colloidal particles, so that the hard sphere interaction between the particles are modified.

We have derived a spherically symmetrical interparticle potential between a pair of the polymer-grafted colloidal particles, using numerical self-consistent field (SCF) calculation. Our SCF results have shown that the pair interaction potential approaches, with some parameter sets, to repulsive step potential containing no attraction~\cite{Norizoe:2005}. This potential was approximated by square-step potential with a rigid core,
\begin{alignat*}{2}
 & \phi (r) = \infty              & \qquad & r < \sigma_1,  \\
 & \phi (r) = \epsilon_0 \; (>0) &        & \sigma_1 < r < \sigma_2,  \\
 & \phi (r) = 0                   &        & \sigma_2 < r,
\end{alignat*}
where $r$ denotes the distance between the centers of the pair of the particles, $\sigma_1$ denotes the diameter of the colloids, and $\sigma_2$ is the diameter of the outer core formed by the polymer brush. A positive constant $\epsilon_0$ represents the repulsion originated from the grafted polymers.

Utilizing molecular Monte Carlo simulation of the particles interacting via $\phi (r)$, we studied the phase behavior of the colloidal system in 2 dimensions. At low temperature and high density, the particles, which have no attractive interaction, start to self-assemble and finally align in strings. We have called this effect ``frogspawn effect" on account of the characteristic particle configuration. The string-like assembly is observed in small regions of the temperature and the density. The width of the potential step, $\sigma_2 / \sigma_1$, where the string-like assembly is found is limited to an interval $1.7 \lessapprox \sigma_2 / \sigma_1 \lessapprox 2.8$, \textit{i.e.} in the vicinity of $\sigma_2 / \sigma_1 = 2.0$. At this value $\sigma_2 / \sigma_1 = 2.0$, when the particles are closely packed on a straight line, both the hard inner cores, diameter $\sigma_1$, and the soft outer cores, diameter $\sigma_2$, of the particles simultaneously pinch and confine each other. This means that the system tends to keep the linear, \textit{i.e.} the string-like, assembly and that the region of $\sigma_2 / \sigma_1$ of the string-like assembly ranges around $\sigma_2 / \sigma_1 = 2.0$.

We also discovered that the string-like assembly indicates a similarity to percolation transition and critical phenomena. The average string length diverges around the region where the melting transition line and the percolation transition line cross. The Fisher exponent, a critical exponent for this percolation transition, $\tau = 1.9$ is kept at any step width $\sigma_2 / \sigma_1$.

In the canonical ensemble ($NVT$-ensemble), the globally-isotropic string-like assembly was found as a metastable structure, \textit{i.e.} an amorphous solid~\cite{Norizoe:2005,2010:NorizoeMuPTLetterArXiv,2011:NorizoeMuPTFullArXiv,2011:NorizoeFrogNVT2DArXiv,Norizoe:2003April,Norizoe:2003November,Master'sThesis,Malescio:2003,Malescio:2004}, whereas the globally-anisotropic defect-free string-like assembly was discovered via 3-reservoirs method as the equilibrated structure~\cite{2010:NorizoeMuPTLetterArXiv,2011:NorizoeMuPTFullArXiv}.

Malescio and Pellicane also simulated the same system at $\sigma_2 / \sigma_1 = 2.0$ and 2.5 and found dimers, trimers, and other various structures~\cite{Malescio:2003,Malescio:2004}. The 2-dimensional system has recently been studied in experiments, where the string-like assembly has actually been confirmed~\cite{Osterman:2007}. These various assemblies were also found in the system at other step widths~\cite{Glaser:2007}, $\sigma_2 / \sigma_1$, and in a system composed of particles interacting via continuous repulsive potential~\cite{Camp:2003} similar to $\phi (r)$.

On the other hand, in 3 dimensions, a glass transition has been observed in the same model system~\cite{Fomin:2008}. A large number of ground states of the same model system have been discovered at zero temperature via genetic algorithms both in 2 and 3 dimensions~\cite{Fornleitner:2008,Pauschenwein:2008}. However, the string-like assembly in 3 dimensions was beyond scope of these recent works. Despite a variety of findings in 2 dimensions shown above, the 3-dimensional systems have been studied far less than the 2-dimensional systems.

In the present work, we simulate our model system and study the phase behavior at finite $T$ in 3 dimensions~\cite{Norizoe:2005}. Simulation methods are given in section~\ref{sec:SimulationMethods}. Simulation results at $\sigma_2 / \sigma_1 = 2.0$ are discussed in section~\ref{sec:SimulationResultsAtStepWidth2.0}. At this step width, the string-like assembly tends to be stabilized~\cite{Norizoe:2005,2011:NorizoeFrogNVT2DArXiv}. Therefore, at this $\sigma_2 / \sigma_1$, we check whether the string-like assembly also appears in 3 dimensions. Simulation results at other step widths are presented in section~\ref{sec:SimulationResultsAtVariousStepWidths}.

\section{Simulation methods}
\label{sec:SimulationMethods}
Monte Carlo simulation in the canonical ensemble ($NVT$-ensemble) via the standard Metropolis algorithm~\cite{ComputerSimulationOfLiquids,Frenkel:UnderstandingMolecularSimulation2002} is performed in 3 dimensions, where $N$, $V$, and $T$ denote the number of particles, the system volume, and the temperature respectively. Mersenne Twister is chosen as a random number generator for our simulation~\cite{MersenneTwister1,MersenneTwister2,MersenneTwister3}. The particles are, in the initial state, arranged on homogeneous fcc lattices in a cubic system box with periodic boundary conditions. In one simulation step, a particle is picked at random and given a uniform random trial displacement within a cube of $0.4 \sigma_1$ long in each direction. A Monte Carlo step (MCS) is defined as $N$ trial moves, during which each particle is chosen for the trial displacement once on average. After $1.0 \times 10^6$ MCS, by which the system relaxes to the equilibrium state except at low temperature and high density, we acquire data every $10^4$ MCS and get 100 independent samples of particle configurations. The number of particles is fixed at $N=4000$.

$\sigma_1$ and $\epsilon_0$ are taken as the unit length and the unit energy respectively. Dimensionless temperature is defined as $k_B T/\epsilon_0$, where $k_B T$ denotes thermal energy. The close-packed volume of the outer cores of the particles is denoted by $V_0$ ($V_0 = N {\sigma_2}^3 \sqrt2 /2$ in 3 dimensions). Dimensionless volume is defined as $V/V_0$.

\section{Simulation results at $\sigma_2 / \sigma_1 = 2.0$}
\label{sec:SimulationResultsAtStepWidth2.0}
Here we discuss simulation results at $\sigma_2 / \sigma_1 = 2.0$, for which the string-like assembly was observed in the 2-dimensional case.

\subsection{Average string length at $\sigma_2 / \sigma_1 = 2.0$}
\label{subsec:AverageStringLengthAtStepWidth2.0}
\begin{figure*}[!tb]
	\centering
	\includegraphics[clip]{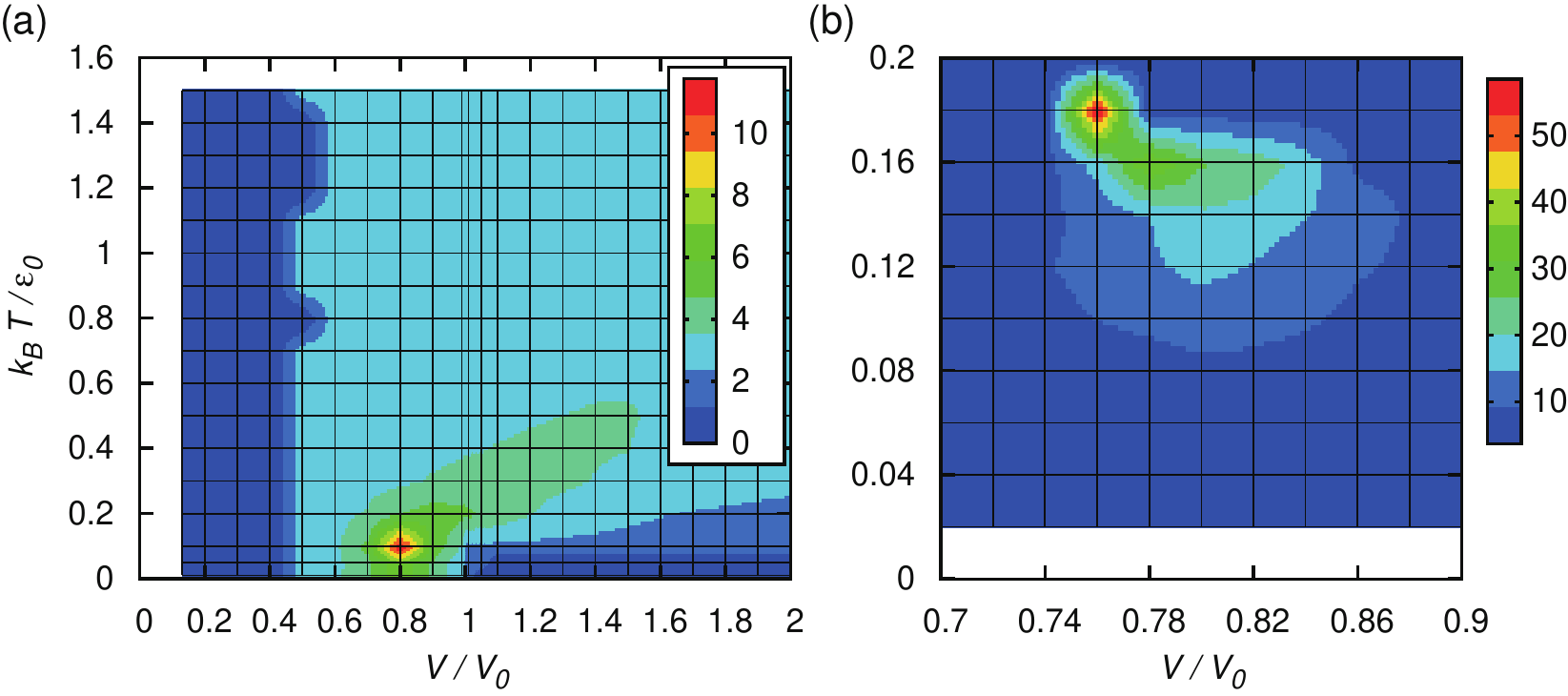}
	\caption{The average string length at $\sigma_2 / \sigma_1 = 2.0$. (b) is the same graph as (a), taken around a peak of (a) with a fine resolution. Blue regions represent short strings and red ones long strings. Grids denote the points where the simulation is performed. Due to statistical errors, small humps are found in a region of $V/V_0 = 0.5$ and $k_B T/\epsilon_0 \ge 0.8$.}
	\label{fig:Frog3DNVTS20AndFine-StringAverageLength}
\end{figure*}

The average string length at $\sigma_2 / \sigma_1 = 2.0$ in 3 dimensions is given in Fig.~\ref{fig:Frog3DNVTS20AndFine-StringAverageLength}. The average string length has a sharp peak at $V/V_0 \cong 0.76$ and $k_B T/\epsilon_0 \cong 0.18$, which means a divergence of the string length. Outside the vicinity of this peak, the long strings are absent. The position of the peak corresponds to the peak in the 2-dimensional system~\cite{Norizoe:2005}, \text{i.e.} $V/V_0 \cong 0.64$ and $k_B T/\epsilon_0 \cong 0.12$.

\begin{figure*}[!tb]
\centering
\includegraphics[clip]{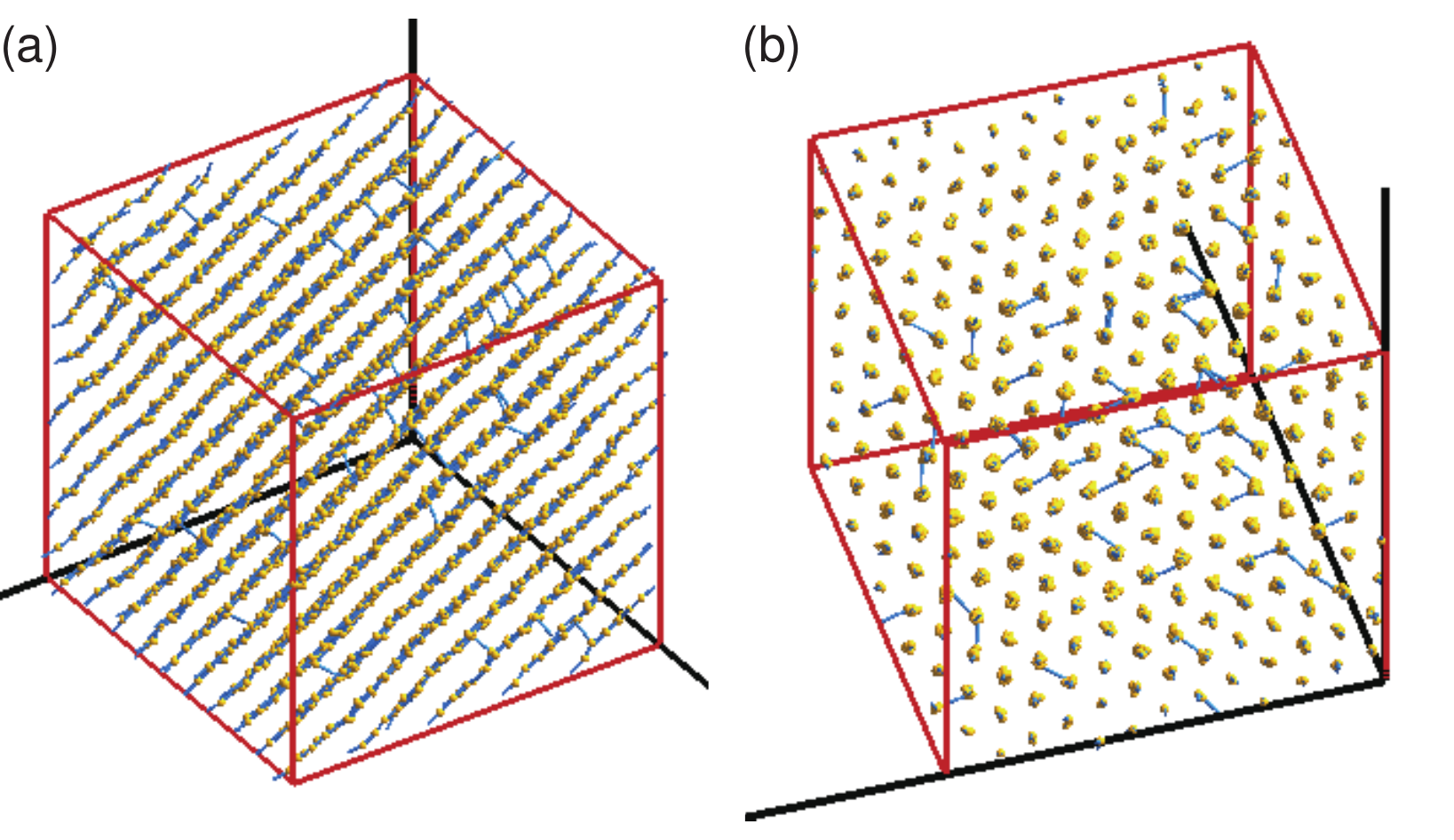}
\caption{Snapshots of the system at $\sigma_2 / \sigma_1 = 2.0$, $V/V_0 = 0.76$, $k_B T/\epsilon_0 = 0.18$, and $1.99 \times 10^{6}$ MCS. Yellow spheres represent the centers of the particles and blue lines denote networks of overlaps between the particles. (a): taken in the lateral direction of the strings. (b): top view of the strings.}
\label{fig:Frog3DNVTS20FineV076T018_001990000MCSLateralTop}
\end{figure*}

Snapshots of the system taken at the peak are shown in Fig.~\ref{fig:Frog3DNVTS20FineV076T018_001990000MCSLateralTop}. These snapshots indicate that the long strings form hexagonally-arranged cylindrical structures, which is similar to cylinder phase typically observed in diblock-copolymers.

These results presented in this section demonstrate that long strings and the string-like assembly are also observed in 3 dimensions, at $\sigma_2 / \sigma_1 = 2.0$.

\subsection{Percolation phenomena at $\sigma_2 / \sigma_1 = 2.0$}
\label{subsec:PercolationPhenomenaAtStepWidth2.0}
\begin{figure*}[!tb]
	\centering
	\includegraphics[clip]{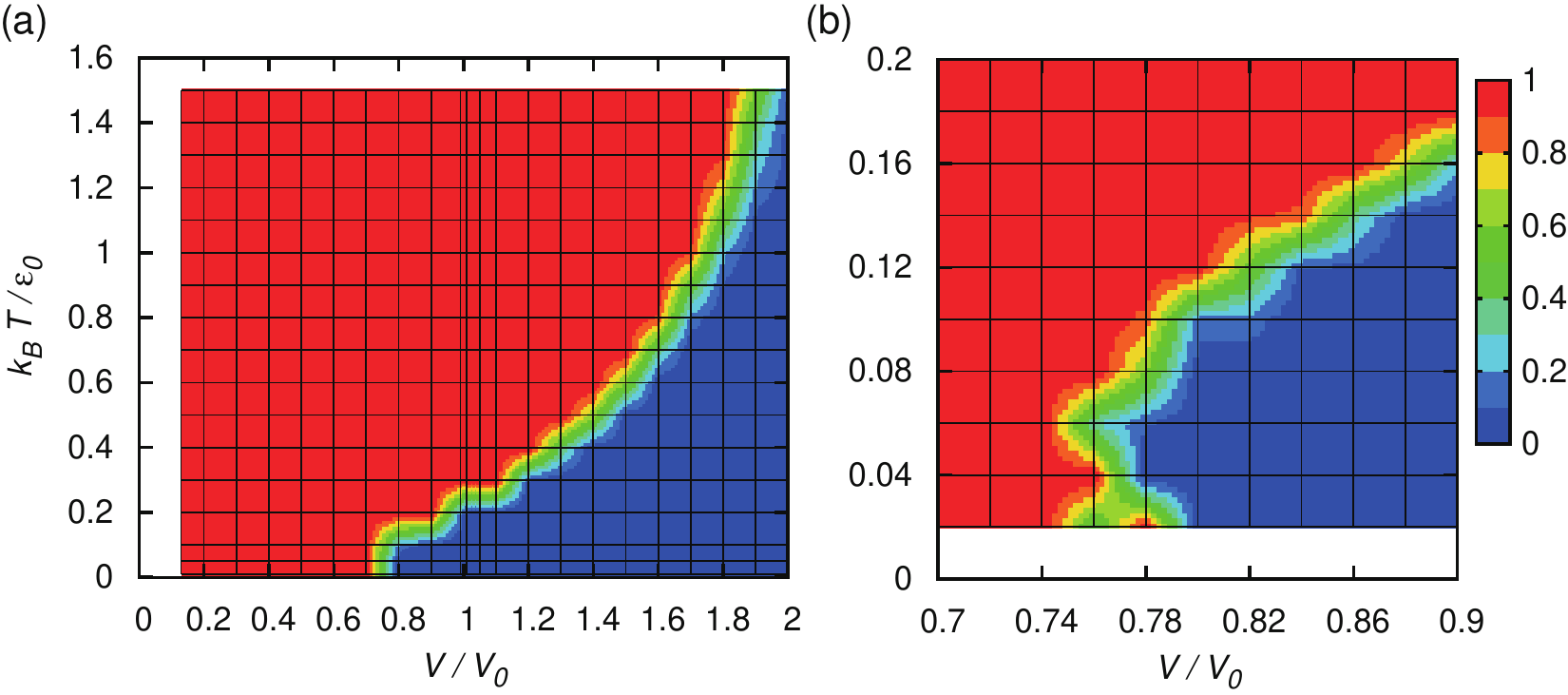}
	\caption{Occurrence probability of percolated clusters for the same system as Fig.~\ref{fig:Frog3DNVTS20AndFine-StringAverageLength}. $\sigma_2 / \sigma_1 = 2.0$. (b) is the same graph as (a), taken with a fine resolution. Blue regions represent non-percolated phase and red ones percolated phase.}
	\label{fig:Frog3DNVTS20AndFine-PercolationTotal}
\end{figure*}

The occurrence probability of a percolated cluster at each parameter set, $k_B T/\epsilon_0$ and $V/V_0$, is given in Fig.~\ref{fig:Frog3DNVTS20AndFine-PercolationTotal}. A sharp boundary, \textit{i.e.} percolation transition line, separates percolation and non-percolation regions. Compared with the results in 2 dimensions~\cite{Norizoe:2005,2011:NorizoeFrogNVT2DArXiv}, the percolated phase extends to high $V/V_0$, \textit{i.e.} low density, regions. Since many degrees of freedom in high dimensions provide a large number of paths from one side of the system box to the opposite side, percolation threshold is lower in 3 dimensions than in 2 dimensions.

\begin{figure}[!tb]
	\centering
	\includegraphics[clip]{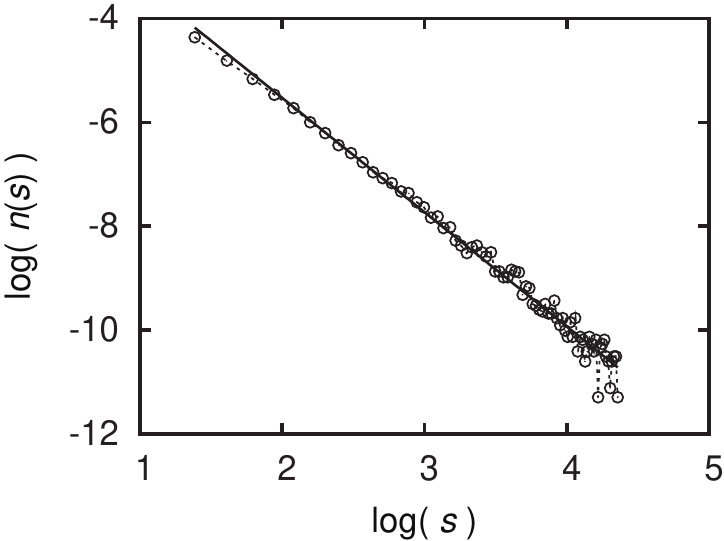}
	\caption{Cluster size distribution, $n(s) = M(s) / N $, at $\sigma _2 / \sigma _1 = 2.0$, $V / V_0 = 1.38$, and $k_B T / \epsilon_0 = 0.5$. $s$ denotes the cluster size and $M(s)$ is the number of clusters with the size $s$ found in the system. This parameter set is located on the percolation transition line in Fig.~\ref{fig:Frog3DNVTS20AndFine-PercolationTotal}. $\log (s)$-$\log (n(s))$ graph is shown. A solid black line shows a linear fitting of these data by $y = -\tau * x + b$ with $\tau = 2.2$ and $b = -1.1$.}
	\label{fig:Frog3DNVTS20V138T05-ClusterSizeDistribution}
\end{figure}

Here we analyze, along the percolation transition line in Fig.~\ref{fig:Frog3DNVTS20AndFine-PercolationTotal}, the cluster size distribution, $n(s)$,
where $s$ denotes the cluster size. An example of $n(s)$ on the percolation transition line is presented in Fig.~\ref{fig:Frog3DNVTS20V138T05-ClusterSizeDistribution}. This graph shows a relation, $n(s) \propto s^{-\tau}$, with $\tau = 2.2$.
This $\tau$ means the Fisher exponent, a critical exponent, of the percolation transition of our system. This power law and the Fisher exponent $\tau = 2.2$ are kept along the percolation transition line. This value, $\tau = 2.2$, in 3 dimensions is slightly higher than $\tau = 1.9$ in 2 dimensions~\cite{Norizoe:2005,2011:NorizoeFrogNVT2DArXiv}. Divergence of the string length in Fig.~\ref{fig:Frog3DNVTS20AndFine-StringAverageLength}, similar to critical phenomena, is found around the percolation transition line. This result is the same as the 2-dimensional system.

\subsection{Thermodynamic phase diagram at $\sigma_2 / \sigma_1 = 2.0$}
\label{subsec:ThermodynamicPhaseDiagramAtStepWidth2.0}
Here we construct a thermodynamic phase diagram at $\sigma_2 / \sigma_1 = 2.0$ in 3 dimensions.

\begin{figure}[!htb]
	\centering
	\includegraphics[clip]{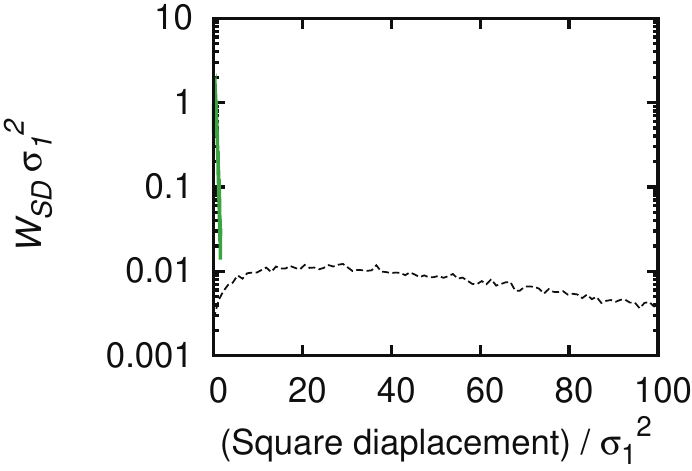}
	\caption{The probability density of the square diaplacement of the particles, $W_{\text{SD}}$, during $1 \times 10^5$ MCS at $\sigma_2 / \sigma_1 = 2.0$ and $k_B T / \epsilon_0 = 0.01$. The result at $V / V_0 = 1.4$ is plotted by a green solid line and $V / V_0 = 1.5$ by a black broken line. The number of analyzed particle configurations, \textit{i.e.} the number of samples, is equal to 9.}
	\label{fig:Frog3DNVTS20V14V15T001-MSDProbabilityDensityTotal}
\end{figure}

In a region of $V / V_0 > 1.0$ and extremely low temperature, our system is regarded as a system composed of hard particles with diamger $\sigma_2$, \textit{i.e.} the outer cores~\cite{Norizoe:2005,2011:NorizoeFrogNVT2DArXiv}. Crystalization of hard particle systems, which is called Alder transition, occurs at $V / V_0 \approx 1.5$ in 3 dimensions~\cite{Alder1960}. Calculating the probability density of the square displacement of the particles in low temperature regions $k_B T / \epsilon_0 \le 0.1$ at $V / V_0 = 1.4$ and 1.5, we confirm this Alder transition of the outer cores of our colloidal particles. The results, \textit{e.g.} presented in Fig.~\ref{fig:Frog3DNVTS20V14V15T001-MSDProbabilityDensityTotal}, show a significant change in this probability density between $V / V_0 = 1.4$ and 1.5. Clearly, this change demonstrates the Alder transition.

In other regions of parameter space for $V / V_0$ and $k_B T / \epsilon_0$, we construct a thermodynamic phase diagram, utilizing the mean-square displacement of the particles (MSD) during $99 \times 10^4$ MCS. Due to the Alder transition of the outer cores, MSD at $V / V_0 = 1.5$ and $k_B T / \epsilon_0 = 0.01$ is set as the criterion to determine whether the system is in a fluid phase or in a solid phase. In the regions of the parameter space where MSD is smaller than this criterion, the system is considered a solid. On the other hand, when MSD is larger than the criterion, the system is a fluid or a coexisting phase of both the phases.

\begin{figure}[!tb]
	\centering
	\includegraphics[clip]{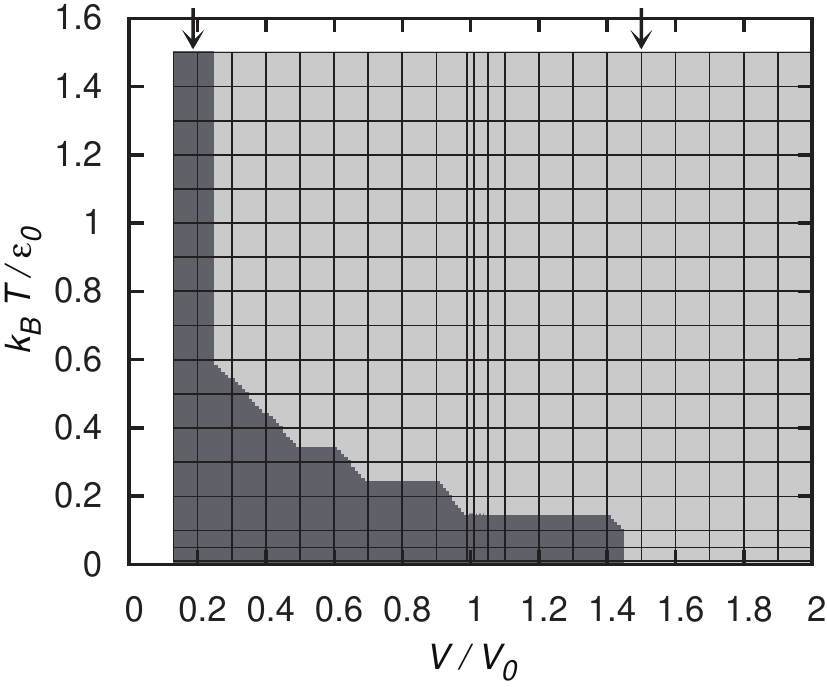}
	\caption{Thermodynamic phase diagram at $\sigma_2 / \sigma_1 = 2.0$. Dark grey regions represent solid phases and light grey fluid or coexistence of both the phases. Alder transition points of the inner and the outer cores are pointed by arrows above the panel.}
	\label{fig:Frog3DNVTS20-ThermodynamicPhaseDiagram}
\end{figure}

The constructed thermodynamic phase diagram is shown in Fig.~\ref{fig:Frog3DNVTS20-ThermodynamicPhaseDiagram}. At high $k_B T / \epsilon_0$ Alder transition of the inner cores of the colloids occurs, whereas Alder transition of the outer cores is not observed. At low $k_B T / \epsilon_0$ only Alder transition of outer cores is observed. At intermediate $k_B T / \epsilon_0$, the solid melts when not only $V/V_0$ but also $k_B T / \epsilon_0$ increases. The region of the string-like assembly and the percolation transition, discussed in section~\ref{subsec:PercolationPhenomenaAtStepWidth2.0}, is located in the middle of the solid phase far from the two Alder transitions, \textit{i.e.} crystalization. In addition to the simulation results in 2 dimensions, the present results in 3 dimensions show that the average string length diverges around the region where the melting transition line and the percolation transition line cross~\cite{Norizoe:2005,2011:NorizoeFrogNVT2DArXiv}.

These results are qualitatively consistent with the results in 2 dimensions~\cite{Norizoe:2005,2011:NorizoeFrogNVT2DArXiv}.

\section{Simulation results at various step widths}
\label{sec:SimulationResultsAtVariousStepWidths}
We have confirmed the existence of the long strings and the string-like assembly in 3 dimensions at $\sigma_2 / \sigma_1 = 2.0$. Here, simulating the system at various step widths, we determine a range of $\sigma_2 / \sigma_1$ in 3 dimensions, where the string-like assembly is observed.

We have discovered~\cite{Norizoe:2005,2011:NorizoeFrogNVT2DArXiv} that the string-like assembly is observed in the vicinity of the step width $\sigma_2 / \sigma_1 = 2.0$. At $\sigma_2 / \sigma_1 = 2.0$, when the particles are closely and linearly packed on a straight line, both the inner cores and the outer cores of the particles simultaneously pinch and confine each other~\cite{2011:NorizoeFrogNVT2DArXiv}. This means that, in this particle configuration, the inner core is pinched by the nearest neighbors and that the outer core by the next nearest neighbors. Therefore, the string-like assembly is stable in the direction of the straight line. When these straight strings are closely arranged in the same direction, the string-like assembly is also stable in the lateral direction since an additional energy cost is required to move and force one particle of a string into the adjacent string. Thus, the system tends to keep the string-like assembly at $\sigma_2 / \sigma_1 = 2.0$. This results in the region of the string-like assembly ranging around $\sigma_2 / \sigma_1 = 2.0$. Our previous simulation results for 2-dimensional systems~\cite{Norizoe:2005,2011:NorizoeFrogNVT2DArXiv} have indicated such a range in $1.7 \lessapprox \sigma_2 / \sigma_1 \lessapprox 2.8$. On the other hand, in 3 dimensions, this interval of $\sigma_2 / \sigma_1$ decreases since the additional degrees of freedom for the 3-dimensional systems reduce the effect of pinch and confinement of the particles arranged along the straight line~\cite{2011:NorizoeFrogNVT2DArXiv}. We simulate the 3-dimensional system and confirm this result.

Percolation phenomena and the thermodynamic phase behavior of the system at various step widths are also studied in the present section.

\subsection{Average string length at various step widths}
\label{subsec:AverageStringLengthAtVariousStepWidths}
\begin{figure*}[!p]
	\centering
	\includegraphics[clip]{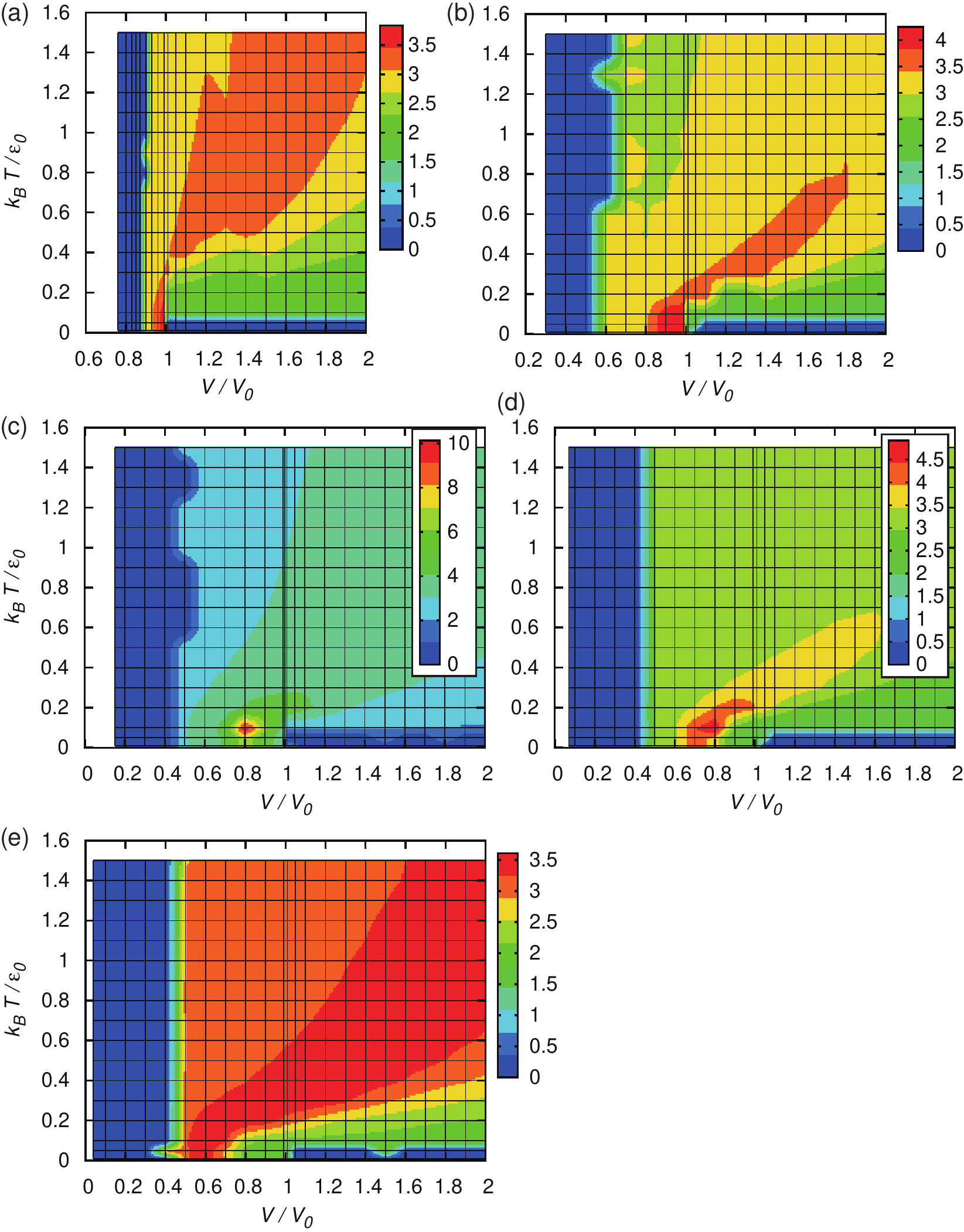}
	\caption{The average string length for $\sigma_2 / \sigma_1 =$ (a): 1.1, (b): 1.5, (c): 1.9, (d): 2.5, and (e): 3.0. Red regions represent long strings and blue regions denote short strings. The average string length at $\sigma_2 / \sigma_1 = 2.0$ is given in Fig.~\ref{fig:Frog3DNVTS20AndFine-StringAverageLength}.}
	\label{fig:Frog3DNVTS11S15S19S25S30-StringAverageLength}
\end{figure*}

\begin{figure}[!tb]
	\centering
	\includegraphics[clip]{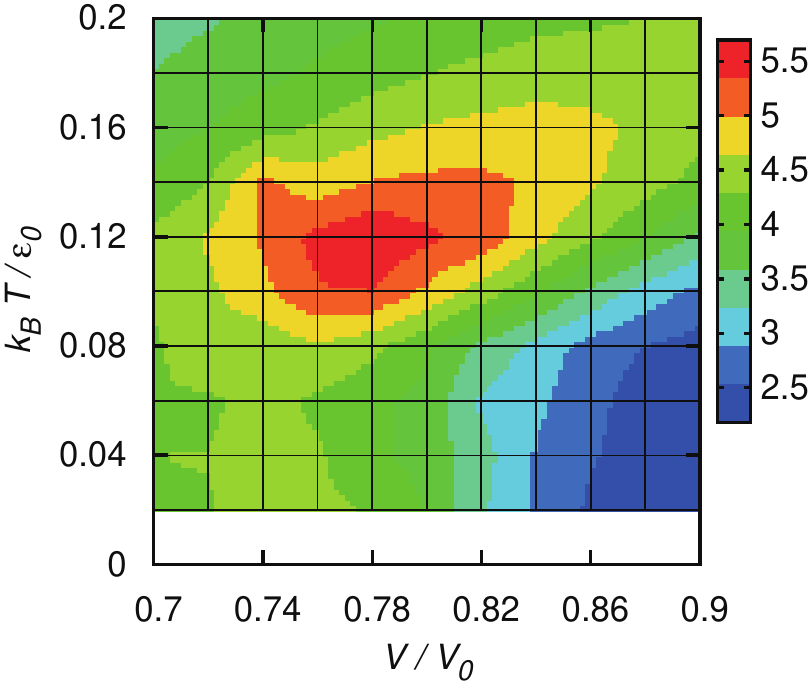}
	\caption{The average string length at $\sigma_2 / \sigma_1 = 2.5$. The same graph as Fig.~\ref{fig:Frog3DNVTS11S15S19S25S30-StringAverageLength}(d), but with higher resolution. The long strings and the divergence of the string length are not observed at this $\sigma_2 / \sigma_1$.}
	\label{fig:Frog3DNVTS25Fine-StringAverageLength}
\end{figure}

The average string length at $\sigma_2 / \sigma_1 = 1.1$, 1.5, 1.9, 2.5, and 3.0 are given in Fig.~\ref{fig:Frog3DNVTS11S15S19S25S30-StringAverageLength}. Long strings and the divergence of the string length are not observed at these values of $\sigma_2 / \sigma_1$. The same quantity as Fig.~\ref{fig:Frog3DNVTS11S15S19S25S30-StringAverageLength}(d), but with high resolution, is shown in Fig.~\ref{fig:Frog3DNVTS25Fine-StringAverageLength}, which illustrates that this result is independent of the resolution of the data. The interval of the step width where the string-like assembly is observed, $1.7 \lessapprox \sigma_2 / \sigma_1 \lessapprox 2.8$ in 2 dimensions~\cite{Norizoe:2005,2011:NorizoeFrogNVT2DArXiv}, is significantly reduced for the present 3-dimensional case. The long strings and the divergence of the string length are found only in the close vicinity of $\sigma_2 / \sigma_1 = 2.0$ in our simulation. This result is consistent with the discussion given above in the present section.

\subsection{Percolation phenomena and thermodynamic phase diagrams at various step widths}
\label{subsec:PercolationPhenomenaAndThermodynamicPhaseDiagramsAtVariousStepWidths}
\begin{figure*}[!p]
	\centering
	\includegraphics[clip]{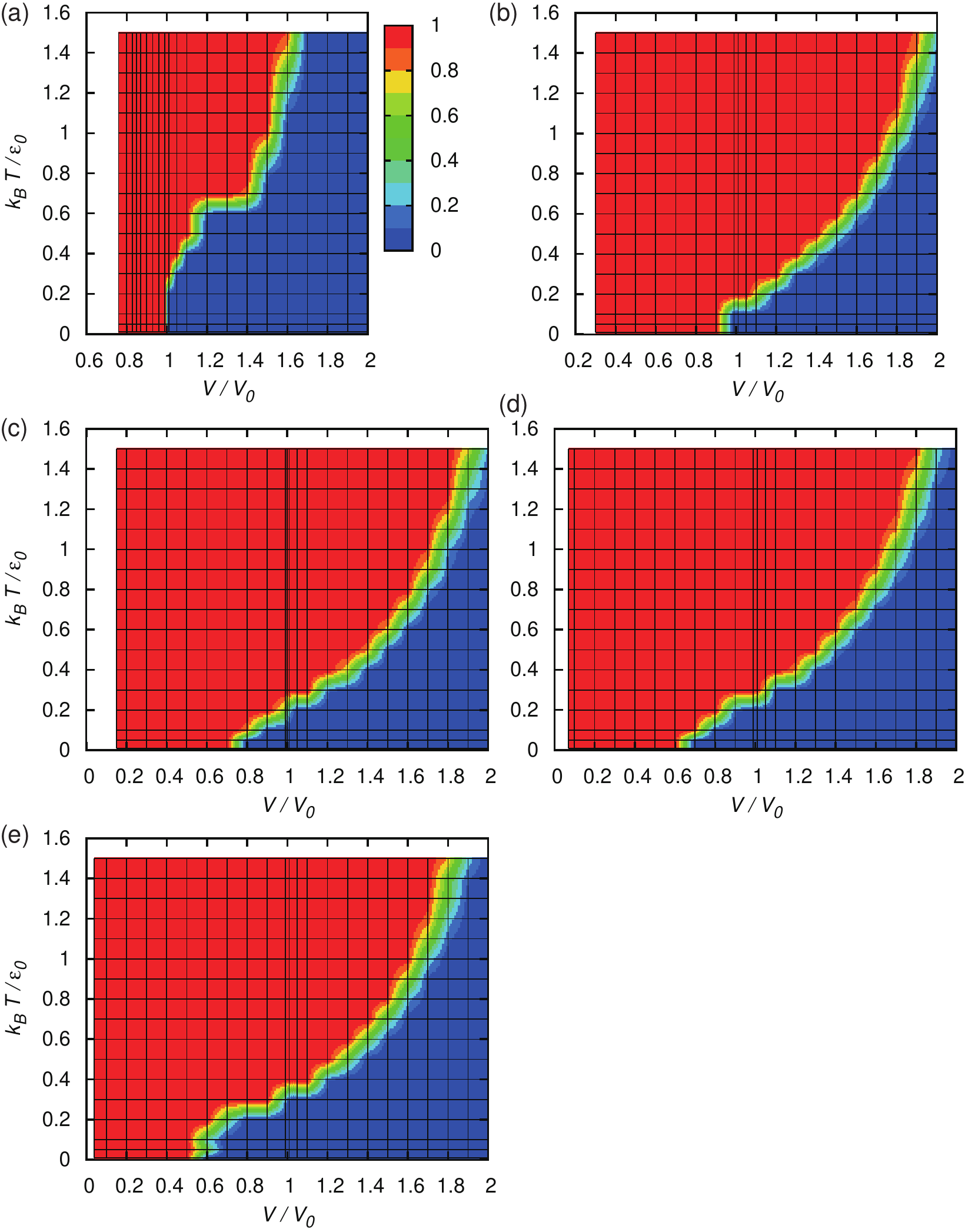}
	\caption{Occurrence probability of percolated clusters for $\sigma_2 / \sigma_1 =$ (a): 1.1, (b): 1.5, (c): 1.9, (d): 2.5, and (e): 3.0. The corresponding occurrence probability of percolated clusters for $\sigma_2 / \sigma_1 = 2.0$ is given in Fig.~\ref{fig:Frog3DNVTS20AndFine-PercolationTotal}.}
	\label{fig:Frog3DNVTS11S15S19S25S30-PercolationTotal}
\end{figure*}

\begin{figure*}[!p]
	\centering
	\includegraphics[clip]{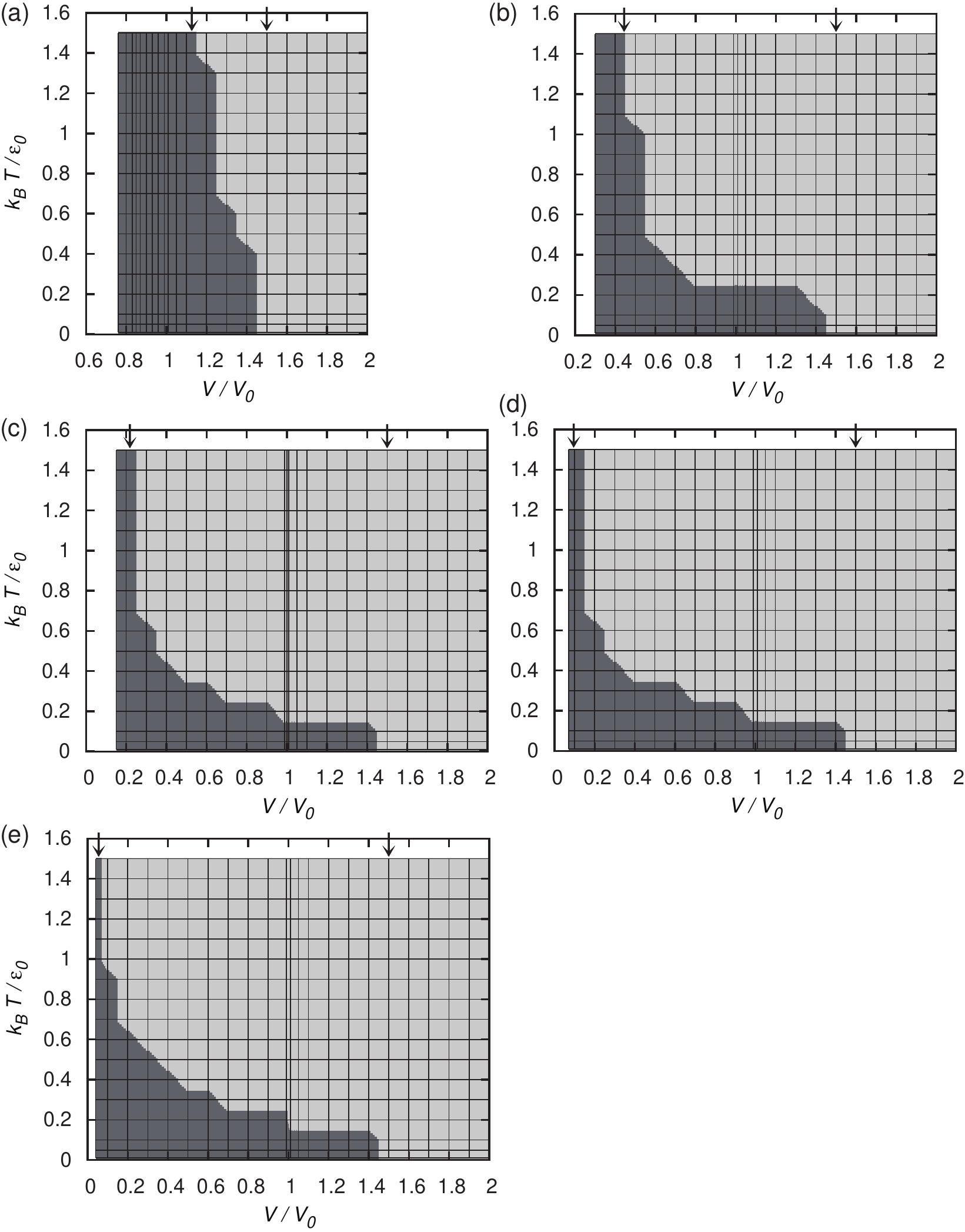}
	\caption{Thermodynamic phase diagrams for $\sigma_2 / \sigma_1 =$ (a): 1.1, (b): 1.5, (c): 1.9, (d): 2.5, and (e): 3.0. The corresponding thermodynamic phase diagram for $\sigma_2 / \sigma_1 = 2.0$ is given in Fig.~\ref{fig:Frog3DNVTS20-ThermodynamicPhaseDiagram}. Dark grey regions represent solid phases and light grey fluid or coexistence of both the phases. Alder transition points of the inner and the outer cores are pointed by arrows above the panel.}
	\label{fig:Frog3DNVTS11S15S19S25S30-ThermodynamicPhaseDiagram}
\end{figure*}

The occurrence probability of percolated clusters and thermodynamic phase diagrams at $\sigma_2 / \sigma_1 = 1.1$, 1.5, 1.9, 2.5, and 3.0 are shown in Figs.~\ref{fig:Frog3DNVTS11S15S19S25S30-PercolationTotal} and \ref{fig:Frog3DNVTS11S15S19S25S30-ThermodynamicPhaseDiagram}. These results are qualitatively consistent with the results at $\sigma_2 / \sigma_1 = 2.0$, given in section~\ref{sec:SimulationResultsAtStepWidth2.0}. The Fisher exponent, $\tau = 2.2$, of the percolation transition of our system is also unchanged at these values of $\sigma_2 / \sigma_1$. These results indicate that the discussion on the results at $\sigma_2 / \sigma_1 = 2.0$ also apply to the results at other values of $\sigma_2 / \sigma_1$.

\section{Conclusions}
\label{sec:Conclusions}
Using particle Monte Carlo simulation, we have studied the phase behavior of 3-dimensional systems composed of the particles interacting via the square-step repulsive potential, $\phi (r)$. The string-like assembly has been confirmed in 3 dimensions, in a solid phase. The long strings compose hexagonally-arranged cylindrical structures, which has a similarity with cylinder phase of diblock-copolymers. The string-like assembly is related to the percolation and the critical phenomena, as was also observed in 2-dimensional systems~\cite{Norizoe:2005,2011:NorizoeFrogNVT2DArXiv}.

At $\sigma_2 / \sigma_1 = 2.0$, when the particles are closely and linearly packed on a straight line, both the inner cores and the outer cores of the particles simultaneously pinch and confine each other. This determines the optimum value of the interval of the step width around $\sigma_2 / \sigma_1 = 2.0$, where the string-like assembly is observed. We have found that the corresponding interval for 2-dimansional systems~\cite{Norizoe:2005,2011:NorizoeFrogNVT2DArXiv} is $1.7 \lessapprox \sigma_2 / \sigma_1 \lessapprox 2.8$, which becomes significantly narrow in the case of 3-dimensional systems. This is originated from the additional degrees of freedom in 3-dimensional space, which reduce the effect of confinement of the particles arranged along the strings.

The Fisher exponent, a critical exponent for the percolation transition, of the system is $\tau = 2.2$, in 3 dimensions, at any step width $\sigma_2 / \sigma_1$. This is slightly higher than $\tau = 1.9$ observed in 2 dimensions~\cite{Norizoe:2005,2011:NorizoeFrogNVT2DArXiv}.

\begin{acknowledgments}
The authors wish to thank Professors Komajiro Niizeki and Andrei V. Zvelindovsky, who gave us helpful suggestions and discussions.
This work is partially supported by a grant-in-aid for science from the Ministry of Education, Culture, Sports, Science, and Technology, Japan.
\end{acknowledgments}


%

\end{document}